\documentclass{article}
\usepackage{cite}
\usepackage{amsfonts}
\usepackage{amssymb}
\usepackage{amsmath}
\usepackage{graphicx}
\usepackage{epstopdf}
\usepackage{slashed}
\usepackage{setspace}
\usepackage{rotating}

\numberwithin{equation}{section}

\usepackage{color}
\definecolor{grn}{rgb}{.0,0.60,0.0}
\definecolor{rd}{rgb}{.60,0.0,0.0}
\definecolor{blk}{rgb}{.0,.0,.0}

\def\tQ{\tilde{Q}}

\def\half{{1\over 2}}

\newcommand{\reef}[1]{(\ref{#1})}

\newcommand{\cn}{{\cal N}}
\newcommand{\cm}{{\cal M}}
\newcommand{\cl}{{\cal L}}

\newcommand{\be}{\begin{equation}}
\newcommand{\ee}{\end{equation}}

\def\be{\begin{equation}}
\def\ee{\end{equation}}
\def\bea{\begin{eqnarray}}
\def\eea{\end{eqnarray}}
\def\ba{\begin{array}}
\def\ea{\end{array}}
\def\bd{\begin{displaymath}}
\def\ed{\end{displaymath}}

\def\a{\alpha}
\def\b{\beta}
\def\c{\gamma}
\def\d{\delta}
\def\td{\tilde \delta}
\def\e{\epsilon}           % Also, \varepsilon
\def\ve{\varepsilon}
\def\g{\gamma}
\def\h{\eta}

\def\l{\lambda}
\def\tl{\tilde \lambda}
\def\m{\mu}
\def\n{\nu}
  
\def\r{\rho}                                     %     \varrho
\def\s{\sigma}                                   %     \varsigma
\def\bs{\bar \sigma}

\def\rmi{\textrm{i}}

\def\pa{\partial}                              % curly d
\def\dg{\sp\dagger} % hermitian conjugate
\def\>{\rangle} %right angle
\def\<{\langle} %left angle
\def\Dsl{D \hskip-.6em \raise1pt\hbox{$ / $ } }
\def\to{\rightarrow}

\def\pa{\partial}

\def\lab{\label}

\newcommand{\lra}{\leftrightarrow}

\def\da{{\dot\alpha}}
\def\db{{\dot\beta}}
\def\dg{{\dot\gamma}}
\def\dd{{\dot\delta}}

\def\tQ{\tilde{Q}}

\def\tve{\tilde{\ve}}

\begin{document}

\setstretch{1.05}

\begin{titlepage}
\begin{flushright}
MIT-CTP-4199 
\end{flushright}
\vspace{1cm}

\begin{center}
{\Large\bf The   $D^{2k}R^4$ Invariants of $\cn=8$~Supergravity}  \\
\vspace{1cm}
{\bf 
Daniel Z.~Freedman${}^{a,b}$ and Erik Tonni$^{b}$} \\

\vspace{0.7cm}

{${}^{a}${\it Department of Mathematics}}\\
{${}^{b}${\it Center for Theoretical Physics}}\\
         {\it Massachusetts Institute of Technology}\\
         {\it 77 Massachusetts Avenue}\\
         {\it Cambridge, MA 02139, USA}\\[5mm]
\end{center}

\vskip .3truecm
\begin{abstract}  
The existence of a linearized SUSY invariant for $\cn=8$ supergravity whose gravitational components are usually called $R^4$ was established long ago by on-shell superspace arguments.  Superspace and string theory methods  have also established analogous higher dimensional  $D^{2k}R^4$ invariants.  However,  very little is known about the SUSY completions of these operators which involve other fields of the theory. In this paper we find the detailed component expansion of the  linearized $R^4$ invariant starting from the corresponding superamplitude
which generates all component matrix elements of the operator.  It is then quite straightforward to extend results to the entire set of $D^{2k}R^4$ operators.

\end{abstract}
\end{titlepage}

%%%%%%%%%%%%%%%
%%%%%%%%%
%\setstretch{0.8}
%\tableofcontents
%\setstretch{1.05}
%%%%%%%%%
%%%%%%%%%%%%%%%

\newpage
\newpage
\setcounter{equation}{0}
\section{Introduction}

The structure of the $\cn = 8$ supergravity theory has been much studied since the theory was first formulated \cite{cremmer, dwdzf}.  It is the maximal supergravity theory in four-dimensional spacetime and appears to have favorable ultraviolet properties \cite{dixon}.   In this paper we are concerned with integral supersymmetric invariants of higher dimension than the classical Lagrangian,  invariants of the
form $\int d^4x \sqrt{-g} (D^{2k} R^4+ \dots)$ where $R$ indicates the Riemann tensor and the dots are terms involving lower spin fields of the theory.  On-shell superspace \cite{brink} techniques have shown that there exist such invariants (with at least linearized on-shell $\cn=8$ SUSY),  but  very little is known about  the lower spin terms.  The purpose of this paper is to determine the SUSY completions in detail.  On-shell SUSY means that the spacetime integral of the variation of the operators vanish  when all component fields satisfy their classical equations of motion.

The particle multiplet of $\cn=8$ SUSY is self-conjugate.  
This feature can be  combined with the spinor-helicity formalism  to obtain compact superamplitudes or generating functions \cite{bef} which encode individual scattering amplitudes.  Tree-level amplitudes are rational functions of
the spinor brackets $\<i\,j\>,~[k\,l]$ formed from external line spinors.   It was emphasized in \cite{efk} that the leading matrix elements of local operators are \emph{polynomials} in the spinor brackets which are strongly constrained by the overall dimension and helicities of the external 
particles.  It was further shown in \cite{efk} that if  the full  superamplitude is a polynomial, it then  corresponds  to an (at least)  linearized supersymmetric operator.
This is one way to show that gravitational operators  $D^{2k}R^4$  have a linearized SUSY completion (for specific distributions of derivatives and index contractions that are implicitly  determined by the matrix elements).  

We will take these methods one step further and derive the specific forms of the SUSY completions of  $D^{2k}R^4$ from the very simple expressions given in \cite{efk} for their superamplitudes.   
The 4-point matrix elements of these operators are all MHV.  
Therefore we need only consider MHV superamplitudes which are  
16th order Grassmann polynomials  of the form \cite{Witten:2003nn}
\be \lab{mhvsamp}
\cm_4^{\rm MHV} =
\d^{(16)}\bigg(\sum_{i=1}^4 | i\> \eta_{ia}\bigg) \frac{m_4(--++)}{\<12\>^8}\,.
\ee
The 4-point graviton matrix element  $m_4(--++)$ determines the full set of component matrix elements related to it by linearized SUSY.  The
 matrix element for any desired set of four external particles is  obtained by applying  a specific Grassmann derivative of order 16, see  \cite{bef}.   
 
 The 4-point matrix element of  the operator $R^4$ is the spinor monomial
 $m_4(--++) = \<12\>^4[34]^4$, so the generating function for its SUSY completion becomes
 \be \lab{r4samp}
\cm^0 =
\d^{(16)}\bigg(\sum_{i=1}^4 | i\> \eta_{ia}\bigg)  \frac{[34]^4}{\<12\>^4}\,.
\ee
Despite the apparent singular denominator, this generating function is a polynomial. To see this,  one can use the explicit form of the Grassmann $\d$-function
\be \lab{delfn}
\d^{(16)}\bigg(\sum_{i=1}^4 | i\> \eta_{ia}\bigg) 
= \frac{1}{256} \prod_{a=1}^8\bigg( \sum_{i,j =1}^4 \<i j\>\h_{ia}\h_{ja}\bigg)\,.
\ee
Each term of the polynomial is a product of eight angle brackets.  Momentum conservation can be used to derive the equality (up to a sign) $[34]/\<12\> =  [ij]/\<kl\>$ for any desired set of four distinct labels $i,j,k,l$.  In this way all potentially singular factors in the denominator of \reef{r4samp} are canceled by factors in the numerator.  The generating function   $\cm_0$ is also totally Bose symmetric. 

We return later to discuss the higher dimensional operators $D^{2k}R^4$.  For the moment it suffices to say that their superamplitudes are given by
\be \lab{d2ksamp}
\cm^k = P_k(s,t,u)\cm^0\,,
\ee
where $P_k(s,t,u)$ is a totally symmetric $k$th order polynomial in the Mandelstam variables.
Specific polynomials are listed in Table 1 of \cite{efk} and earlier in the  string literature \cite{Greenstu}.

The SUSY invariants $D^{2k}R^n+\ldots$ are candidate counterterms to cancel ultraviolet divergences in  $n$-point S-matrix elements at loop order
$L= k+n-1$.  Explicit calculations  have shown that 4-point matrix elements
are finite through $L=4$ loop order \cite{Bern4}.  The $R^4$ invariant was already excluded as an actual counterterm by earlier 3-loop calculations \cite{Bern3}.  There is no $D^2 R^4$ invariant simply because the only available symmetric polynomial is $P_1(s,t,u) = s+t+u$, which vanishes. Nevertheless, the 4-loop calculations are valuable, because the possible divergences are studied in dimensional regularization. Results through 4-loop order suggest  that the critical dimension at which the  S-matrix first diverges is the same in both $\cn=8$ SG and in the dimensionally continued $\cn=4$ SYM theory.  A 5-loop calculation would provide a critical test of this conjectured property  \cite{Bjornsson:2010wm}. 

Recently, information  from the $\a'$ expansion \cite{Stieberger} of string theory amplitudes has been used to show that the $R^4$ invariant cannot appear as an actual counterterm
\cite{EK}  because its nonlinear SUSY completion produces amplitudes which violate the low energy theorems of spontaneously broken $E_{7(7)}$ symmetry.  Similar argumentation shows \cite{befkms} that the potential 5- and 6-loop $D^4R^4$ and $D^6R^4$ counterms are also absent.  These facts were suspected from earlier suggestive arguments \cite{grv, Bossard}.   
Attention therefore focuses on the 7-loop $D^8R^4$ invariant as the possible lowest divergence of the $\cn=8$ theory.

\section{Particle states and fields in $\cn=8$ SG}

The 256 particle states of $\cn =8$  transform in antisymmetric products of the fundamental representation of the SU(8)  R-symmetry group.  The tensor rank $r$ and helicity $h$ are related by $2h=4-r$.  Thus the annihilation operators for helicity states of the graviton,  8 gravitini, 28 graviphotons, 56 graviphotini, and 70 scalars may be listed as:
\be \lab{annih}
A^+,  ~A^a,~A^{ab}, A^{abc}, \ldots, A^{abcdefg},~A^{abcdefgh}\,.
\ee
The last two entries describe  the 8 helicity -3/2  gravitini and helicity -2 graviton.  The upper index notation is the most convenient one to  extract  individual amplitudes from the generating function. However,  one can always use the SU(8) Levi-Civita tensor to lower indices.  For example,  one can equally well use the lower index operators defined via 
$ A^{abcde} =  \tfrac{1}{3!} \e^{abcdefgh}A_{fgh}$ to describe helicity -1/2  graviphotini.\footnote{The inverse relation  $A_{a_1\ldots a_r}  = (-)^r\e_{a_1\ldots a_8} A^{a_{r+1}\ldots a_8}/(8-r)!$ contains a - sign for fermions.}

The chiral supercharges  $Q^a \equiv -\ve^\a Q^a_\a$ and $\tilde{Q}_a\equiv \tilde{\ve}_{\dot \a} Q_a^{\dot \a}$ act on particle states of on-shell matrix elements.    It is convenient to include the (anti-commuting) SUSY parameters $\ve^\a$ and   $\tilde{\ve}_{\dot \a}$ in these definitions, in which
$\a$, $\da$ are standard Weyl spinor indices.  The supercharge $\tilde{Q}_a$ raises helicity by 1/2 unit, and $Q^a$ lowers helicity.
We also represent the SUSY parameters as  angle and square spinors, i.e.  $\ve\to 
|\ve],~~\tilde{\ve} \to |\tve\>$.  In general we (try to) adhere to the spinor-helicity conventions of \cite{bef}. Some useful formulas are collected in Appendix \ref{app spinors}.   

The action of the supercharges on the various $A^{..}$ operators is determined by the helicity properties  and SU(8) covariance.   Thus
$[\tQ_a,\, A^+]=0$, because  the helicity +2 of the operator $A^+$ is maximal in the multiplet.   Some other examples of the supercharge commutation relation are  (with spinors $|p\>,~|p]$ labeled
by the particle momentum)
\be \label{examples}
[\tQ_a,A^b] =\<\tve p\> \d^b_a A\, ,  \qquad [\tQ_a,A^{bc}] = \<\tve p\>(\d^b_aA^c - \d^c_aA^b)\, , \qquad
[Q^a,A^b] = [p \ve] A^{ab}\,.
\ee
Complete details are given in (2.6) of \cite{bef} .    

The SUSY algebra must be satisfied, which means that
\be\label{alg}
[[\tQ_a,Q^b],A^{\cdots}]   =   -\d^b_a\<\tve|p|\ve] A^{\cdots}
\ee
on all operators $A^{\cdots}$.   As an example of an algebra check, we give
\bea \label{algck}
[[\tQ_a,Q^b],A^{c}] &=& [\tQ_a,[Q^b,A^c]] - [Q^b,[\tQ_a,A^c]]\\ 
&=& [\tQ_a,A^{bc}][p\ve] - \d^c_a\<\tve p\>[Q^b,A^+]\\
&=&- \<\tve|p|\ve] \big( (\d^b_a A^c - \d^c_aA^b)  + \d^c_a A^b\big) = -\d^b_a \<\tve|p|\ve]A^c\,.
\eea
Try it, it's fun.  The more indices, the more fun.

We need a set of field operators $\Phi^{\cdots}(x)$, one for each $A^{\cdots}$ and SUSY variations $\d\Phi^{\cdots}(x)$ which are faithful to the structure of $[\tQ_a,A^{\cdots}]$ and $[Q^a,A^{\cdots}]$ and the algebra \reef{alg}.  In the free field limit,   each $\Phi^{\cdots}(x)$ must communicate to the particle of the same helicity carried by the corresponding  $A^{\cdots}$.   The fields which do this are the symmetric spinor fields of the Penrose-Newman formalism \cite{penrose}.

\section{Gauge-invariant multi-spinor fields}
\label{section gauge-inv}

It is well known that the electromagnetic gauge field strength can be expressed in the spinor formalism;  see  \cite{penrose} for details.
The relation between tensor and spinor components is
\be  \lab{emspin}
F_{\m\n} = \frac{1}{4}(\bs_\m)^{\da\a} (\bs_\n)^{\db\b}\big(\e_{\da\db} F_{\a\b} + \e_{\a\b} F_{\da\db}\big)\,.
\ee
The physical content of the six real components of $F_{\m\n}$ is captured in the three complex components of the symmetric $F_{\a\b}$ (which is the conjugate of  $F_{\da\db}$.)   Only the anti-symmetry of $F_{\m\n}$ is needed to derive the representation \reef{emspin}.   One can then show that
the gauge field Bianchi identity and the Maxwell field equation imply that
$\pa^{\dg\a}F_{\a\b}\,=0\,.$   Since the spinor space is two-dimensional, this is equivalent to the symmetry relation
\be \lab{symderiv}
\pa_{\a \dd}F_{\b\g} = \pa_{\b \dd}F_{\a\g}\,.
\ee
This relation and its extensions to the gravitino and gravitational fields will be very useful for us.  

A similar spinor decomposition holds for the on-shell curvature tensor. We refer readers to \cite{vannwu} and \cite{penrose} for the derivation and simply state that after imposing symmetry properties, the algebraic Bianchi identity, and the Einstein equations $R_{\m\n}=0$,  one finds that the physical information in the ten independent components of $R_{\m\n\r\s}$ can be reexpressed in terms of the five complex components of the fourth rank symmetric spinor   $R_{\a\b\g\d}$ (and its dotted spinor conjugate).
The original differential Bianchi identity can then be expressed  
 \cite{penrose} as $\pa^{\dot{\r}\a}R_{\a\b\g\d}\,=0\,,$  which is equivalent to the useful symmetry property
 \be \lab{gravsym}
 \pa_{\s\dot{\r}}R_{\a\b\g\d} = \pa_{\a\dot{\r}}R_{\s\b\g\d} \,.
 \ee

The  (4-component spinor)  gauge-invariant gravitino field strength
$\psi_{\m\n} = \pa_\m\psi_\n-\pa_\n\psi_\m$ satisfies the simplified  
\cite{Bible}  Rarita-Schwinger equation
\be \lab{rssimp}
\g^\m\psi_{\m\n}=0\,.
\ee
There are $6\times 2=12$ complex components in each chiral projection of
$\psi_{\m\n}$  and the field equation \reef{rssimp} imposes  $4\times 2=8$ conditions on these.  
It should be no surprise that the physical information in the 4 independent on-shell  components can be rexpressed in terms of a third rank symmetric
spinor.   However, we have not found the needed discussion in the literature, so we try to present a self-contained treatment.  We use the chiral
projectors $P_{\pm} = \half (1 \pm \g_5)$, and write the $P_+$ projection of
$\psi_{\m\n}$ as the 2-component spinor 
\be \lab{gino1}  
\psi_{\m\n\dg} = \frac{1}{4}(\bar{\s}_\m)^{\da\a} (\bar{\s}_\n)^{\db\b}\big(\e_{\da\db} \psi_{\a\b\dg} + \e_{\a\b} \psi_{\da\db\dg}\big)\,.
\ee
So far we have only used the anti-symmetry in $\m\n$.  Next we note that
$\psi_{\m\n\dg}$ satisfies the (projected) equation of motion
\be \lab{rseom} 
(\s^\m)_{\g\dg}\psi_{\m\n}{}^{\dg}\,=\,0\,.
\ee
Upon substitution of \reef{gino1}  and use of detailed properties \cite{bef} of the $\s^\mu$ and $\bar{\s}_\mu$,  one can show that the equation of motion implies that $\psi_{\a\b\dg}=0$ and that $\psi_{\da\db\dg}$ is a totally symmetric spinor.  

The on-shell third rank spinors $\psi_{\a\b\g}$ and $\psi_{\da\db\dg}$ are the 2-component parts of the 4-component tri-spinor
$ (\g^{\m\n})\psi_{\m\n}$.  We  will use standard Dirac algebra to establish
the desired symmetry property of derivatives of  $\psi_{\a\b\g}$ and $\psi_{\da\db\dg}$.   We insert the Dirac operator and consider the tri-spinor
$(\g^{\m\n}\g^\r)\,\pa_\r\psi_{\m\n}$.  The 2-component parts of this quantity
are $\pa^{\dot{\r}\a}\psi_{\a\b\g}$  and its conjugate. We will show that the entire 4-component tri-spinor vanishes.  This follows from
\be
(\g^{\m\n}\g^\r)\,\pa_\r\psi_{\m\n} = (\g^{\m\n\r} +\g^\m \eta^{\n\r} -\g^\n\eta^{\m\r})\,\pa_\r\psi_{\m\n}\,,
\ee
if we note two facts.  First the cyclic combination $\pa_\r\psi_{\m\n}  +\pa_\m \psi_{\n\r} +\pa_\n\psi_{\r\m} $  vanishes; it is a Bianchi-like identity satisfied by $\psi_{\m\n}$.   Second, the contracted first derivative  $\pa^\m \psi_{\m\n}$ also vanishes if the conventional on-shell plane expansion is used. Thus we learn that $\pa^{\dot{\r}\a}\psi_{\a\b\g}=0$;  the symmetric derivative property then follows as above:
\be \lab{symder3/2}
 \pa_{\s\dot{\r}}\psi_{\a\b\g} = \pa_{\a\dot{\r}}\psi_{\s\b\g} \,.
 \ee

The symmetric spinor fields of rank $s$ have simple plane-wave expansions in which the "modes" are simply the $2s$-fold product of spinors $\l_\a(p)$ or $\tl_\da(p)$.  For example the expansions of the third rank gravitino fields are  (with conventional integration measure  $dp \equiv d^3 p /[(2\pi)^3 2p^0]$)
\bea \lab{ginoexp}
\psi_{\a\b\g}(x) &=& \int dp\, \l_\a(p) \l_\b(p) \l_\g(p)\Big(  e^{\rmi p \cdot x} B(p)  + 
 e^{-\rmi p \cdot x} D^\dagger(p) \Big)\,,\\
\psi_{\da\db\dg}(x) &=& \int dp\, \tl_\da(p) \tl_\db(p) \tl_\dg(p)\Big(  e^{\rmi p \cdot x} D(p)  + 
 e^{-\rmi p \cdot x} B^\dagger(p) \Big)\;.
\eea 
 Here $B(p)$ and $D(p)$ are annihilation operators for positive and negative helicity particles, respectively.  Their adjoints are creation operators.
  
The plane wave expansions of the fields are tabulated below, including
appropriate SU(8) indices.  We always omit the eight indices 12345678 on graviton
operators, and thus use the notation  $A^+$ and $A_-$ as the annihilation operators for gravitons.   The tabulation begins with the operator $ R_{\da \db \dg \dd} $,  which contains the annihilator $A^+$, and then extends downward in helicity.
   Scalar fields satisfy the SU(8) self-duality condition  $\phi^{abcd} = \frac{1}{4!}\e^{abcdefgh}\phi_{efgh}.$    

 \begin{eqnarray} \lab{pwexps}
R_{\da \db \dg \dd}  
&=&  \int dp
\, \tilde{\l}_{\da}(p)   \tilde{\l}_{\db}(p)   \tilde{\l}_{\dg}(p)   \tilde{\l}_{\dd}(p)  
\Big(A^+(p) e^{\rmi p \cdot x} + A^+(p)^\ast e^{-\rmi p \cdot x} \Big)\,,
 \nonumber \\
\psi^a_{\da \db \dg} 
&=&  \int dp
\, \tilde{\l}_{\da}(p)   \tilde{\l}_{\db}(p)  \tilde{\l}_{\dg}(p) 
\Big( A^a(p) e^{\rmi p \cdot x} + A^a(p)^\ast e^{-\rmi p \cdot x} \Big)\,,
 \nonumber  \\
 F^{ab}_{\da \db} 
&=&  \int dp
\, \tilde{\l}_{\da}(p)  \tilde{\l}_{\db}(p)   
\Big( A^{ab}(p) e^{\rmi p \cdot x} + A^{ab}(p)^\ast e^{-\rmi p \cdot x} \Big)\,,
  \nonumber  \\
 \chi^{abc}_{\da} 
&=&  \int dp
\, \tilde{\l}_{\da}(p)  
\Big( A^{abc}(p) e^{\rmi p \cdot x} + A^{abc}(p)^\ast e^{-\rmi p \cdot x}  \Big)\,,
  \nonumber  \\
  \phi^{abcd} 
&=&  \int dp
\Big( A^{abcd}(p) e^{\rmi p \cdot x} + A^{abcd}(p)^\ast e^{-\rmi p \cdot x}  \Big)\,,
  \nonumber  \\
   \phi_{abcd} 
&=&  \int dp
\Big( A_{abcd}(p) e^{\rmi p \cdot x} + A_{abcd}(p)^\ast e^{-\rmi p \cdot x}  \Big)\,,
  \nonumber \\
  \chi^\a_{abc}
&=&  \int dp
\, \l^{\a}(p)  
\Big( A_{abc}(p) e^{\rmi p \cdot x} + A_{abc}(p)^\ast e^{-\rmi  p \cdot x}  \Big)\,,
  \nonumber \\
  F^{\a\b}_{ab}
&=&  \int dp
\, \l^{\a}(p)  \l^{\b}(p)  
\Big( A_{ab}(p) e^{\rmi p \cdot x} + A_{ab}(p)^\ast e^{-\rmi p \cdot x}  \Big)\,,
  \nonumber \\
   \psi^{\a\b\g}_a
&=&  \int dp
\, \l^{\a}(p)   \l^{\b}(p)  \l^{\g}(p)  
\Big( A_{a}(p) e^{\rmi  p \cdot x} + A_{a}(p)^\ast e^{-\rmi p \cdot x}  \Big)\,,
  \nonumber \\
    R^{\a\b\g\d}
&=&  \int dp
\, \l^{\a}(p)   \l^{\b}(p)  \l^{\g}(p)   \l^{\d}(p)  
\Big( A_-(p) e^{\rmi p \cdot x} + A_-(p)^\ast e^{-\rmi p \cdot x}  \Big)\,.
 \end{eqnarray}
The superscript  $^\ast$ is a formal notation for creation operators,  which are precisely defined as the adjoints of the annihilation operator for the corresponding anti-particle of opposite helicity.  Thus, for example
 \be
A_-(p)^\ast =A^{+}(p)^\dagger ,
\hspace{9mm}
 A_{a}(p)^\ast = A^{a}(p)^\dagger,
\hspace{9mm}
A_{ab}(p)^\ast =A^{ab}(p)^\dagger,
\hspace{9mm}
 A^{a}(p)^\ast = A_{a}(p)^\dagger\,.
\ee

\section{SUSY transformations of multi-spinor fields} 

We now use the SUSY transformation rules  for annihilators discussed in Sec. 2 to derive the transformations for on-shell multi-spinor fields.  We treat the graviphoton fields  $F^{bc}_{\da\db}$ and $F^{\a\b}_{ab}$ explicitly
and thereby infer the general structure. 
Let's compute:
\bea \lab{delaF}
\td_a F^{bc}_{\da\db}  &=&  \int dp\, \tl_\da(p) \tl_\db(p)\Big( e^{\rmi p\cdot x} [\tQ_a, A^{bc}]+ \ldots \Big)
 \nonumber \\
  &=&   \int dp \,\tl_\da(p) \tl_\db(p) \Big( e^{\rmi p\cdot x} \<\tve p\>(\d^b_a A^c -\d^c_a A^b) + \dots \Big)
   \nonumber \\
&=&  -\, \tve^\dg \big( \d^b_a \psi_{\da\db\dg}^c -  \d^c_a \psi_{\da\db\dg}^b \big)\,,
\eea
and similarly e.g. for the gravitino
\bea
\d^a\psi^b_{\da\db\dg} &=& \int  dp\,\tl_\da(p) \tl_\db(p) \tl_\dg(p) 
\Big(e^{\rmi p\cdot x} [Q^a, A^{b}]+ \ldots \Big)
 \nonumber \\ 
&=& \ve^\a \int dp \, p_{\a\da} \tl_\db(p) \tl_\dg(p) \Big(e^{\rmi p\cdot x} A^{ab} + \ldots \Big)
 \nonumber \\
&=&  -\rmi\,\ve^\a \pa_{\a\da} F^{ab}_{\db\dg}\,,
\eea 
where we used $\l_\a(p) \tl_\da(p)=-p_{\a\da}$. Note that the right side of the last equation is symmetric in $\da\db\dg$ because of the symmetric derivative property \reef{symderiv}.  \\
The SUSY transformations of the full set of multi-spinor fields are
\begin{equation}
\begin{array}{lcl}
\tilde{\delta}_{a} R_{\dot{\alpha}\dot{\beta}\dot{\gamma}\dot{\delta}}  
\,=\, 0
& &
\d^a R_{\dot{\alpha}\dot{\beta}\dot{\gamma}\dot{\delta}} 
\,=\,
-\rmi \,\varepsilon^\sigma \partial_{\sigma\dot{\alpha}} \psi^a_{\dot{\beta}\dot{\gamma}\dot{\delta}}  
\\
%& & \nonumber
%\\
\rule{0pt}{.6cm}
\tilde{\delta}_{a} \psi^b_{\dot{\a}\dot{\b}\dot{\g}}  
\,=\,  -\tilde{\varepsilon}^{\dot{\rho}} \delta_a^b R_{\dot{\rho}\dot{\alpha}\dot{\beta}\dot{\gamma}} 
& &
\d^a  \psi^b_{\dot{\alpha}\dot{\beta}\dot{\gamma}}  
\,=\,
-\rmi\,\varepsilon^\sigma \partial_{\sigma\dot{\alpha}} F_{\dot{\beta}\dot{\gamma}}^{ab}
\\
%& & \nonumber
%\\
\rule{0pt}{.6cm}
\tilde{\delta}_{a} F_{\dot{\alpha}\dot{\beta}}^{bc}  
\,=\, -\tilde{\varepsilon}^{\dot{\rho}} 
\big( \delta_a^b \psi^c_{\dot{\rho}\dot{\alpha}\dot{\beta}} 
-   \delta_a^c \psi^b_{\dot{\rho}\dot{\alpha}\dot{\beta}} \big)
& &
\d^a  F^{bc}_{\dot{\alpha}\dot{\beta}}  
\,=\,
-\rmi\,\varepsilon^\sigma \partial_{\sigma\dot{\alpha}} \chi_{\dot{\beta}}^{abc}
%\\
%& & \nonumber
\\
\rule{0pt}{.6cm}
\tilde{\delta}_{a} \chi_{\dot{\alpha}}^{bcd}  
\,=\, -\tilde{\varepsilon}^{\dot{\rho}} 
\big( \delta_a^b F^{cd}_{\dot{\rho}\dot{\alpha}} 
+\delta_a^c F^{db}_{\dot{\rho}\dot{\alpha}} 
+\delta_a^d F^{bc}_{\dot{\rho}\dot{\alpha}}  \big)
& &
\d^a  \chi^{bcd}_{\dot{\alpha}}  
\,=\,
-\rmi\,\varepsilon^\sigma \partial_{\sigma\dot{\alpha}} \phi^{abcd}
\\
%& & \nonumber
%\\
\rule{0pt}{.6cm}
\tilde{\delta}_{a} \phi^{bcde}  
\,=\, -\tilde{\varepsilon}^{\dot{\rho}} 
\big( \delta_a^b \chi^{cde}_{\dot{\rho}} 
-\delta_a^c \chi^{deb}_{\dot{\rho}} 
+\delta_a^d \chi^{ebc}_{\dot{\rho}} 
-\delta_a^e \chi^{bcd}_{\dot{\rho}} 
 \big)
& &
\d^a  \phi^{bcde} 
\,=\,
-\varepsilon^\sigma \chi_{\sigma}^{abcde}
\\
\rule{0pt}{.4cm}
\tilde{\d}_{a} \phi_{bcde}  
\,=\, -\tilde{\ve}_{\dot{\rho}}  \chi^{\dot{\r}}_{abcde}
& &
\d^a  \phi_{bcde} 
\,=\,
\ve_\s \big( \d^a_b \chi^\s_{cde} - \d^a_c \chi^\s_{deb} + \d^a_d \chi^\s_{ebc} -
\d^a_e \chi^\s_{bcd}  \big)
\\
%& & \nonumber
%\\
\rule{0pt}{.6cm}
\tilde{\delta}_{a} \chi^{\a}_{bcd}  
\,=\,  -\rmi\,\tilde{\ve}_{\dot{\rho}}  \partial^{\dot{\rho}\a} \phi_{abcd}
& &
\d^a \chi^{\a}_{bcd}  
\,=\, -\ve_\sigma \big( 
\d^a_b F^{\sigma\alpha}_{cd}  + \d^a_c F^{\sigma\alpha}_{db} + \d^a_d F^{\sigma\alpha}_{bc} \big)
\\
%& & \nonumber
%\\
\rule{0pt}{.6cm}
\tilde{\delta}_{a} F^{\a\b}_{bc}  
\,=\,  \rmi\,\tilde{\ve}_{\dot{\r}}  \partial^{\dot{\r}\a} \chi^{\b}_{abc}
& &
\d^a  F^{\a\b}_{bc}  
\,=\,
\ve_\s \big(\d^a_b \psi^{\sigma\alpha\beta}_{c} - \d^a_c \psi^{\sigma\alpha\beta}_{b} \big)
\\
%& & \nonumber
%\\
\rule{0pt}{.6cm}
\tilde{\delta}_{a} \psi^{\a\b\g}_{b}  
\,=\, -\rmi\,\tilde{\ve}_{\dot{\rho}} \partial^{\dot{\r}\a} F^{\beta\gamma}_{ab}
& &
\d^a  \psi^{\a\b\g}_{b}  
\,=\,-
\ve_\s \d^a_b R^{\s\a\b\g}
\\
%& & \nonumber
%\\
\rule{0pt}{.6cm}
\tilde{\d}_{a} R^{\alpha\beta\gamma\delta}
\,=\,  \rmi\,\tilde{\ve}_{\dot{\rho}}  \partial^{\dot{\rho}\a}
 \psi^{\beta\gamma\delta}_{a}
& &
\d^a  R^{\alpha\beta\gamma\delta}
\,=\, 0 \,.
\end{array} \label{trafos}
\end{equation}
Given these transformations, it is straightforward to check the SUSY algebra. For instance, we have
\be
\big[\d^a, \tilde{\d}_b\big]\psi^c_{\dot{\a}\dot{\b}\dot{\g}}    \,=\,
\big(\d^a \tilde{\d}_b- \tilde{\d}_b \d^a\big) \psi^c_{\dot{\a}\dot{\b}\dot{\g}}
\,=\,-\rmi\,\d^a_b  \,\ve^\s  \tilde{\ve}^{\dot{\rho}} \pa_{\s\dot{\rho}}  \psi^c_{\dot{\a}\dot{\b}\dot{\g}} \,.
\ee
The symmetric derivative relations of Sec. 3 are crucial in checks of the algebra  and elsewhere.

%%%%%%%%%%%%%%%%%%%%%%%% 
 \section{The method and an example}
 
In our work, the leading 4-point matrix element of a quartic SU(8) invariant Lagrangian $\cl$
is represented as a product of 4 annihilation operators from the list in \reef{annih} acting to the left on the "out" vacuum.   As explained in \cite{bef} each matrix element is obtained from the generating function \reef{mhvsamp}    by applying the 16th order  Grassmann derivative which uniquely corresponds to the upper SU(8) indices carried by the 4 individual particles.  SU(8) symmetry requires that each (non-vanishing) matrix element contain a total of 16 upper indices with each index $a,b=1,2,\dots, 8$ paired.  The 16 indices are distributed among the 4 annihilators, as in the example 
\be \lab{8620exam}
\<{\rm out} |A_-(1)A^{345678}(2) A^{12}(3) A^+(4) \int \cl\,  |{\rm in} \>\,
=\,
\< 8620 | \int \cl\, |{\rm in} \>\,\,.
\ee
 Recall that there are
8 suppressed upper indices $12345678$ on the -ve helicity graviton operator, so the example conforms to  the general rule.   

Our method consists of three steps.  We state the procedure roughly at first and then refine it as needed:\newline
 1.   Obtain the matrix elements of all independent 4-point amplitudes from the generating  function \reef{mhvsamp}.  The evaluation of 16th order derivatives can be recast as the Wick contraction algorithm discussed in Sec. 3.2 of \cite{bef}.   There is a shortcut  which quickly gives any amplitude up to a sign.  In the 21st century, the most efficient method is to use a Mathematica program which automates the calculation of any 16th order derivative and thus any matrix element.  This is the way we will do it.\footnote{We thank Henriette Elvang for sharing her code with us.} \newline
  2.  For each SU(8) independent amplitude, there is a quartic Lagrangian $\cl$ which is an SU(8) invariant contraction of 4 operators from the list in  
\reef{pwexps}.  Each Lagrangian term is unique up to partial integration of the 
spacetime derivatives it contains.  The form of the matrix element in Step 1 tells us how to place these derivatives and contract spinor indices. \newline
3.  The SUSY transformations of the multi-spinor fields
were defined in (\ref{trafos}) to be faithful to the SUSY properties of the corresponding annihilation operators.   It therefore follows that the sum of all SU(8) independent operators constructed in Steps 1 and 2 is a linearized SUSY invariant.

How many independent matrix elements are there?  Let's observe that the upper index pattern of the 4 operators in the example \reef{8620exam}  corresponds to the (8620) partition of the integer 16.  In general there is an
allowed choice of 4 annihilation operators (and therefore an allowed MHV 4-point process) for every partition of 16 of length 4 with maximum summand 8.   There are 33 such partitions.  For some partitions, there is
a unique SU(8) invariant contraction of the 4 fields.  This is the case for the
(8620) partition for which we have the product of fields:
$\dot{R} F^{ab} F_{ab}R$  (we temporarily omit spinor indices and use 
 $\dot{R}$ to indicate the  curvature multi-spinor with dotted indices).  For others, such as the partition (7432), there is more than one, in this case the
 two contractions  $\psi^b \phi^{cdef}\chi_{bcd}F_{ef}$ and $\psi^b \phi^{cdef}\chi_{cde}F_{fb}$.   
 For each invariant monomial there is one (actually more than one)  outgoing 4-particle state  
  which "communicates" to that monomial.  The state used is specified by
 a particular choice of SU(8) quantum numbers.

The matrix elements we work with are strongly constrained by scaling requirements.  First each  of the two   conjugate operators for a spin $s$ particle carries effective scale dimension $s$ in a matrix element.  Thus the total dimension of the 4 operators in  \reef{8620exam} is $2+1+1+2 =6$.   However the  overall dimension of all terms in the component expansion of
$R^4$ must have scaling dimension 8.   The missing two units of dimension are spacetime derivatives, which are automatically supplied by the specific amplitude computed by Grassmann differentiation.  Let's see how this happens.  The $(8620)$ amplitude is
\be  \lab{8620amp}
M^{(8620)} = \<12\>^2\<13\>^2[34]^4 =  \<12\>^2 \<1|3|4]^2 [34]^2.
\ee
The momentum $p_3$ appears twice, so the Lagrangian $\cl^{(0268)}$ will have two derivatives with respect to the field which communicates to the positive helicity graviphoton.  Of course, momentum conservation implies that  $\<1|3|4] = - \<1|2|4]$.   If we use this we get an equivalent alternate form of   $\cl^{(0268)}$ 
in which a derivative is applied to the field which comunicates to the negative helicity graviphoton.  The two forms of the Lagrangian  are related by partial integration.  We should expect no less (and no more)!    Amplitudes are also constrained by the helicity scaling
relation which originates \cite{Witten:2003nn} in the energy dependence of external line spinors and polarization vectors.  For each particle $i=1,2,3,4$ the difference between the number of angle spinors $a_i$ and square spinors $s_i$ is $a_i -s_i=-2h_i$. Indeed, \reef{8620amp} contains the weights
$a_1-s_1 = 4, ~a_2-s_2 = 2,~ a_3-s_3 = -2,~a_4-s_4=-4$, as required.

We need to be more concrete about the correspondence between annihilation operators and the multi-spinor fields they "communicate" with.
The main point is that an annihilation operator such as  $A_{12}(2) =A^{345678}(2)$ for a -ve helicity graviphoton communicates to the negative frequency part of the conjugate field  $F^{12}_{\da\db}$.   The effective Wick contraction is
\be \lab{8620wick1}
 e^{- \rmi p_2\cdot x} \tl_{\da}(2) \tl_{\db}(2)\,.
 \ee
Similarly the annihilation operator $A^{12}(3)$ communicates to the conjugate field $F^{\a\b}_{12}$ with Wick contraction
 \be \lab{8620wick2}
 e^{- \rmi p_3\cdot x} \l^{\a}(3) \l^{\b}(3)\,.
 \ee
 Similar remarks apply to the graviton operators in \reef{8620exam}.
 The detailed form of the matrix element \reef{8620amp} then leads us to  the corresponding Lagrangian term
 \be \lab{0268lag}
\cl^{(0268)}= - \half R_{\da\db\dg\dd}F^{\da\db\,ab}\pa^{\dg\g}\pa^{\dd\d}F^{\a\b}_{ab}R_{\a\b\g\d}\,.
 \ee
 The superscript (0268) is the partition which counts the number of upper SU(8) indices of the fields involved.
 This form of the matrix element is unique except for the possibility of integration by parts.  One can see that upon partial integration, the derivative $\pa_{\g\dg}$ is non-vanishing only when it "hits" the field $F_{\da\db}^{ab}$.  Other terms vanish due to the differential constraints on the multi-spinor fields.   The factor $\half$ above compensates for the double-counting in the sum on $ab$.   The minus sign comes from the product of two factors of $-\rmi$ from the correspondence $p_{\g\dg} \lra -\rmi \pa_{\g\dg}$.
This is the only place a sign can emerge for the $(8620)$ matrix element because other possible sources, such as order of spinor index contractions,
are always paired.  In other cases it is more difficult to determine the sign.
We do our best.

%%%%%%%%%%%%%%
\section{The $R^4$ invariant}
\label{section R4 inv}

Here in more detail are the  steps of the procedure to determine the various operators which contribute to the $R^4$ integral invariant.  
\begin{enumerate}
\item Partitions   $(r_1, r_2, r_3, r_4)$, with  $r_i \ge r_{i+1}$ correspond to the various matrix elements we need.  The complementary partition $(8-r_1,8-r_2,8-r_3,8-r_4)$ describes the operator (or operators) to which the particles of the partition couple.
\item  Write in schematic form the product of fields in the quartic monomials for each partition including all independent invariant contractions of SU(8) indices.  At this stage  only  SU(8) is considered, so spinor indices are omitted. Further, fields for identical particles are treated as independent,  distinguished by the order in which they appear in the monomials. 
One finds 63 algebraically independent SU(8) invariants.
\item  Select numerical values of the SU(8) indices of the particles in the matrix element that couple to each of these monomials. 
Use a computer code or manual method  (it's not hard) to obtain the matrix element from the
generating function \reef{r4samp}.  The result quickly gives the correct coupling of spinor indices and derivatives and thus gives the correct operator up to a multiplicative constant.
\item  Determine this constant precisely by computing the matrix element
$\< A^{..}(1)A^{..}(2)A^{..}(3)A^{..}(4)\int \cl\,\>$ carefully and matching to the result of step 3. 
%It is important to include all Wick contractions when the matrix element contains identical particles.  
\item A special  situation can occur in partitions which contain identical particles.  
Studying these cases, we found that for 10 such partitions  a smaller number of field monomials suffices to generate all matrix elements.   In this way  the number of field monomials needed is reduced to 51.  An example of this reduction is discussed below.  (Note that one must be careful to include all Wick contractions to compute matrix elements.)
  \end{enumerate}

It is interesting to compare the  procedure outlined above   to the combinatoric analysis of (effectively) 4-point amplitudes in $\cn =8$ SG  in the latter part of Sec 4.6 of \cite{efkwi}.  The analysis there confirms the 33 partitions and initial basis of 63 amplitudes.  For each partition, the number of field monomials in our initial list is equal to the Kostka number of the partition.  The Kostka number counts the number of SU(8) singlets in the direct product of the
four SU(8) irreps of the fields in the monomial.  However, as discussed in \cite{efkwi},   the Kostka number can overcount functionally independent amplitudes in partitions containing two (or more) identical particles, and 
the partitions in which we find reductions in the number of field monomials
are partitions in which  functional relations were found in \cite{efkwi}.

We discuss the partition $(7711)$ as an example  of our approach.  There are two independent SU(8) invariant monomials at stage 2 of the procedure  outlined above.  They are 
\be  \lab{7711mons}
L_1 = \psi^b\psi^c \psi_b \psi_c\,,~~~~~~~~L_2 =  \psi^b\psi^c \psi_c \psi_b\,.
\ee
   We then consider the outgoing state $\<\rm{out} | A_1(1) A_2(2) A^1(3) A^2(4)$ for which the generating function \reef{r4samp} gives the matrix element
\be  \lab{1212amp}
\< A_1(1) A_2(2) A^1(3) A^2(4)\,S\, \> = \<12\>^3 s_{23} [34]^3\,.
\ee
This state communicates to both monomials, and we reproduce the correct matrix element from both $L_1$ and $L_2$ if we assign spinor indices, derivatives, and multiplicative constant 
\bea
L_1    &\to & \cl_1=-\tfrac{2}{2}\psi_{\da\db\dg}^b
 \partial_\m  \psi^{\da\db\dg\,c}
\partial^\m  \psi^{\a\b\g}_b
 \psi_{\a\b\g\,c} \\
L_2    &\to & \cl_2=  \tfrac{2}{2}\psi_{\da\db\dg}^b
 \partial_\m  \psi^{\da\db\dg\,c}
  \psi^{\a\b\g}_c
\partial^\m \psi_{\a\b\g \,b} \,.
 \eea
These two operators are identical, so there is a unique operator in the (7711) sector, which we rename $\cl_1 = \cl$.
It is straightforward to verify, with proper attention to Wick contractions and fermion anticommutation, that
\be
\< A_1(1) A_2(2) A^1(3) A^2(4)\int \cl\, \> = \<12\>^3 s_{23} [34]^3.
\ee
One also  obtains the correct matrix elements for two independent external states, namely
\bea
\< A_1(1) A_2(2) A^2(3) A^1(4)\int \cl\, \> &=& \<12\>^3 s_{24} [34]^3\\
\< A_1(1) A_1(2) A^1(3) A^1(4)\int \cl \,\> &=&- \<12\>^3 s_{12} [34]^3\,.
\eea
In the last case there are four Wick contractions and one uses $s_{23} +s_{24}=-s_{12}$
to produce the result above.

We used the method above to determine the 51 independent operators in the list below.  The sum of these operators, integrated over spacetime gives the desired $R^4$ invariant.  The procedure guarantees that the result has linearized supersymmetry.  This means that its $Q^a$ and $\tQ_a$ variations, computed using the transformation rules of  \reef{trafos}, vanish.

It would be useful to verify linearized SUSY to check the signs and other details of the operators in the list.  However,
a complete check  is prohibitively difficult because  many independent field monomials appear in the variation,  and the variation as many as four operators from the list contributes to each monomial.  Furthermore there are many opportunities for sign errors due to incorrect raising and lowering of spinor indices and inattention to anti-commutativity among spinor fields and SUSY parameters.  We have made  two careful SUSY checks  to confirm  the signs of the first 
four terms in the list.   These checks are described in Appendix \ref{app susy}.  A third detailed check was done which confirms the form and coefficient of the operator for the (7711) partition discussed above.
In addition we have made several more checks up to signs which illustrate how independent monomials in the SUSY variations cancel   due to the Schouten relation and momentum conservation.  
  
The component expansion of the $R^4+\ldots$ invariant is the sum of 51 operators in the third column in the tabulation below.  The partition of 16 which specifies the four fields in each operator is given in the first column.  The second column lists the spinor-helicity form of the matrix element of each operator.\footnote{A few matrix elements were computed in \cite{KLR}.}
The fractional coefficients in each monomial are not reduced to lowest terms in order to indicate the origin of various factors.  Factors of 2 in numerators  arise from the Mandelstam formula $s_{ij} = -2p_i\cdot p_j$.
Factors of 3! in denominators cancel the overcounting in contractions of 3 SU(8) indices on one pair of fields.  Factors of 2 in denominators  avoid overcounting of pairs of SU(8) indices or  account for multiple Wick contractions of identical fields.  There are pairs of partitions, such as (8422) and (6640), whose particle states are related by charge conjugation. The corresponding operators are then each other's
adjoints.   The 10 partitions discussed in point 5 above are labeled  by $\circ$  (or $\circ\circ $) to indicate how many SU(8) invariants are redundant.   
\newpage

\be
\begin{array}{lclcl}
\label{LIST}
\<r_1\,r_2\,r_3\,r_4|
 &\hspace{.2cm}& 
 \textrm{\bf matrix element} &\hspace{0cm}& \textrm{\bf monomial }  \mathcal{L}^{(8-r_1\,8-r_2\,8-r_3\,8-r_4)}
 \\ 
\rule{0pt}{.7cm}
%\langle - - + + \rangle  & & 
\<8\,8\,0\,0| &&  \langle 12 \rangle^4 [34]^4
&&
\tfrac{1}{4}R_{\da\db\dg\dd}  R^{\da\db\dg\dd} 
R^{\a\b\g\d}  R_{\a\b\g\d}  
\\
%\langle\, - \, \psi^- \, \psi^+ \,+ \,\rangle  && 
\rule{0pt}{.4cm}
\<8\,7\,1\,0| && 
\langle 12 \rangle^3 \langle 1|3|4]  [34]^3
&&
-\rmi R_{\da\db\dg\dd}  
\psi^{\da\db\dg\,b} 
\partial^{\dd\d}
 \psi^{\a\b\g}_b   R_{\a\b\g\d}  
\\
%\langle\, - \, V^- \, V^+ \,+ \,\rangle  && 
\rule{0pt}{.4cm}
\<8\,6\,2\,0| && 
\langle 12 \rangle^2 \langle 1|3|4]^2  [34]^2
&&
-\tfrac{1}{2}
R_{\da\db\dg\dd}   F^{\da\db\,bc} 
\partial^{\dg\g} \partial^{\dd\d} 
  F_{bc}^{\a\b}  
R_{\a\b\g\d}  
\\
%\langle\, - \, V^- \, \psi^+ \,\psi^+ \,\rangle  && 
\rule{0pt}{.4cm}
\<8\,6\,1\,1| && 
-\langle 12 \rangle^2 \langle 1|2|3]  \< 1|2|4]  [34]^2
&&
-\tfrac{1}{2}
R_{\da\db\dg\dd} 
   \partial^{\dg\g}\partial^{\dd\d} 
F^{\da\db\,bc}   
\psi^{\a\b}_{\hspace{.36cm}\g\,b}      
\psi_{\a\b\d\,c}  
\\
%\langle\, - \, \chi^- \, \chi^+ \,+ \,\rangle  && 
\rule{0pt}{.4cm}
\< 8\,5\,3\,0| && 
\langle 12 \rangle \langle 1|3|4]^3  [34]
&&
 \tfrac{\rmi}{3!} 
 R_{\da\db\dg\dd}  \chi^{\da\,bcd}  
\partial^{\db\b} \partial^{\dg\g} \partial^{\dd\d} 
  \chi^{\a}_{bcd}
 R_{\a\b\g\d}  
\\
%\langle\, - \, \chi^- \, V^+ \, \psi^+ \,\rangle  && 
\rule{0pt}{.4cm}
\< 8\,5\,2\,1| && 
\langle 12 \rangle \langle 1|2|3]  \<1|3|4]^2  [34]
&&
\tfrac{\rmi}{2}
R_{\da\db\dg\dd}  \partial^{\da\a}  \chi^{\dd\,bcd}
  \partial^{\db\b} \partial^{\dg\g} 
   F^{\d}_{\hspace{.14cm}\a\,bc} 
  \psi_{\b\g\d\,d}
\\
%\langle\, - \, \phi \, \phi \,+ \,\rangle  && 
\rule{0pt}{.4cm}
\< 8\,4\,4\,0 | && 
 \langle 1|2|4]^2  \langle 1|3|4]^2
&&
\tfrac{1}{4!}
R_{\da\db\dg\dd}  
\partial^{\da\a} \partial^{\db\b} \phi^{bcde}
\partial^{\dg\g}  \partial^{\dd\d} \phi_{bcde}
R_{\a\b\g\d}  
\\
%\langle\, - \, \phi \, \chi^+ \,\psi^+ \,\rangle  && 
\rule{0pt}{.4cm}
\< 8\,4\,3\,1| && 
\langle 1|2|3] \langle 1|3|4]^3 
&&
-\tfrac{1}{3!} 
R_{\da\db\dg\dd}  \partial^{\da\a}  \phi^{bcde}
\partial^{\db\b} \partial^{\dg\g}
\partial^{\dd\d}  \chi_{\a\,bcd} 
\psi_{\b\g\d\,e}  
\\
%\langle\, - \, \phi \, V^+ \,V^+ \,\rangle  && 
\rule{0pt}{.4cm}
\< 8\,4\,2\,2 | && 
\langle 1|2|3]^2 \langle 1|2|4]^2 
&&
\tfrac{1}{2^3}
R_{\da\db\dg\dd}
\partial^{\da\a} \partial^{\db\b} \partial^{\dg\g}  \partial^{\dd\d} 
\phi^{bcde}
F_{\a\b\,bc} 
 F_{\g\d\,de}  
\\
%\langle\, - \, \chi^+ \, \chi^+ \,V^+ \,\rangle  && 
\rule{0pt}{.4cm}
\<8\,3\,3\,2| && 
\langle 1|2|3] \langle 1|2|4] \langle 1|3|2] \langle 1|3|4]
&&
-\tfrac{1}{3!^2 2^2} 
R_{\da\db\dg\dd}
 \e^{bcdefghi} 
\partial^{\da\a}  \partial^{\db\b} \chi_{\g\,ghi}
\partial^{\dg\g} \partial^{\dd\d}   \chi_{\a\,bcd}  
F_{\b\d\,ef} 
\\
%\langle\, \psi^- \, \psi^- \, V^+ \,+ \,\rangle  &&
\rule{0pt}{.4cm} 
\< 7\,7\,2\,0| && 
-\langle 12 \rangle^2 \langle 1|3|4] \langle 2|3|4] [34]^2
&&
\tfrac{1}{2}\psi_{\da\db\dg}^b \psi^{\da\db\hspace{.14cm}c}_{\hspace{.34cm}\dd} 
\partial^{\dg\g}  \partial^{\dd\d}
  F^{\a\b}_{bc}
R_{\a\b\g\d}  
\\
%\langle\, \psi^- \, \psi^- \, \psi^+ \,\psi^+ \,\rangle  && 
\rule{0pt}{.4cm}
\< 7\,7\,1\,1| \, \circ&& 
\langle 12 \rangle^3 s_{23}   [34]^3
&&
-\tfrac{2}{2}\psi_{\da\db\dg}^b
 \partial_\m  \psi^{\da\db\dg\,c}
\partial^\m  \psi^{\a\b\g}_b
 \psi_{\a\b\g\,c} 
\\
%\langle\, \psi^- \, V^- \, \chi^+ \,+ \,\rangle  && 
\rule{0pt}{.4cm}
\< 7\,6\,3\,0|  && 
\langle 12 \rangle \langle 1|2|4]^2  \langle 2|3|4]   [34]
&&
-\tfrac{\rmi}{2}\psi^b_{\da\db\dg} F^{\dg\hspace{.14cm}cd}_{\hspace{.18cm}\dd}
\partial^{\da\a} \partial^{\db\b} 
\partial^{\dd\d}
\chi^{\g}_{bcd}
R_{\a\b\g\d}  
\\
%\langle\, \psi^- \, V^- \, V^+ \,\psi^+ \,\rangle  && 
\rule{0pt}{.4cm}
\< 7\,6\,2\,1| && 
-\langle 12 \rangle^2 \langle 1|3|4]  s_{23}   [34]
&&
2 \rmi  \psi_{\da\db\dg}^b 
  \partial_\m F^{\da\db\,cd}
\partial^\m \partial^{\dg\g} 
F^{\a\b}_{bc}
\psi_{\a\b\g\,d}  
\\
\rule{0pt}{.4cm}
&& 
\langle 12 \rangle^2 \langle 1|3|4]  s_{24}   [34]
&&
\tfrac{2\rmi}{2}
\psi_{\da\db\dg}^b 
  \partial_\m F^{\da\db\,cd}
 \partial^{\dg\g} 
F^{\a\b}_{cd}
\partial^\m \psi_{\a\b\g\,b}  
\\
%\langle\, \psi^- \, \chi^- \, \phi \,+ \,\rangle  && 
\rule{0pt}{.4cm}
\< 7\,5\,4\,0 | && 
-\langle 1|3|4]^3   \langle 2|3|4]^3 
&&
-\tfrac{1}{3!}\psi_{\da\db\dg}^b \chi_{\dd}^{cde}
\partial^{\da\a} \partial^{\db\b} 
\partial^{\dg\g} \partial^{\dd\d}\phi_{bcde}
R_{\a\b\g\d}  
\\
%\langle\, \psi^- \, \chi^- \, \chi^+ \,\psi^+ \,\rangle  && 
\rule{0pt}{.4cm}
\< 7\,5\,3\,1| && 
\<12\> \<1|3|4]^2 s_{23} [34]
&&
-\tfrac{2}{2}\psi_{\da\db\dg}^b 
\partial_\m   \chi^{\da\,cde}
\partial^\m  \partial^{\db\b}  \partial^{\dg\g} \chi^\a_{bcd}
\psi_{\a\b\g\,e} 
\\
\rule{0pt}{.4cm}
 && 
\<12\> \<1|3|4]^2 s_{24} [34]
&&
\tfrac{2}{3!}\psi_{\da\db\dg}^b 
\partial_\m   \chi^{\da\,cde}
 \partial^{\db\b}  \partial^{\dg\g} \chi^\a_{cde}
\partial^\m \psi_{\a\b\g\,b} 
\\
%\langle\, \psi^- \, \chi^- \, V^+ \,V^+ \,\rangle  && 
\rule{0pt}{.4cm}
\< 7\,5\,2\,2 |\, \circ && 
- \<12\> \<1|3|4] \<1|4|3] s_{23} [34]
&&
\tfrac{2}{2} \psi_{\da\db\dg}^b 
  \partial_\m \chi^{\da\, cde}
  \partial^\m \partial^{\dg\g} F_{\a\b\,bc} 
 \partial^{\db\b}
 F^\a_{\hspace{.18cm}\g\,de} 
\\
%\langle\, \psi^- \, \phi \, \phi \,\psi^+ \,\rangle  && 
\rule{0pt}{.4cm}
\< 7\,4\,4\,1| && 
- \< 1|3|4]^3   s_{23}  
&&
\tfrac{2\rmi}{3!2}\psi_{\da\db\dg}^b 
 \partial_{\m} \phi^{cdef}
\partial^{\da\a} \partial^{\db\b} 
 \partial^{\dg\g}  \partial^{\m}  \phi_{bcde}
\psi_{\a\b\g\,f}  
\\
\rule{0pt}{.4cm}
&& 
\< 1|3|4]^3   s_{24}  
&&
-\tfrac{2\rmi}{4!2} \psi_{\da\db\dg}^b 
 \partial_{\m} \phi^{cdef}
\partial^{\da\a} \partial^{\db\b} 
 \partial^{\dg\g}   \phi_{cdef}
\partial^{\m}  \psi_{\a\b\g\,b} 
\\
%\langle\, \psi^- \, \phi \, \chi^+ \,V^+ \,\rangle  && 
\rule{0pt}{.4cm}
\< 7\,4\,3\,2| && 
\<1|2|3] \<1|3|4]^2  s_{24}  
&&
-\tfrac{2\rmi}{3!}\psi_{\da\db\dg}^b \partial_{\m} 
\partial^{\da\a} \phi^{cdef}\,
  \partial^{\db\b} \partial^{\dg\g}    
\chi_{\a\,cde} 
\partial^{\m}  F_{\b\g\,bf} 
\\
\rule{0pt}{.4cm}
&& 
-\<1|2|3] \<1|3|4]^2  s_{23}  
&&
\tfrac{2\rmi}{2^2}\psi_{\da\db\dg}^b \partial_{\m} 
\partial^{\da\a} \phi^{cdef}\,
\partial^{\m}  \partial^{\db\b} \partial^{\dg\g}    
\chi_{\a\,bcd} 
  F_{\b\g\,ef}  
\\
\rule{0pt}{.4cm}
\< 7\,3\,3\,3| \,\circ&& 
\< 1|2|3] \< 1|3|4]  \< 1|4|2]  s_{23}   
&&
\tfrac{2\rmi}{3!^2 6}\psi_{\da\db\dg}^b
\e^{cdefghij}
\partial_{\m}  \partial^{\da\a}  \chi_{\g\,hij}
  \partial^{\m}  \partial^{\db\b}  \chi_{\a\,bcd} 
  \partial^{\dg\g}  \chi_{\b\,efg}  
 \\
 \rule{0pt}{.4cm}
\< 6\,6\,4\,0| && 
\< 1|3|4]^2 \<2|3|4]^2
&&
\tfrac{1}{2^3}F_{\da\db}^{bc} F_{\dg\dd}^{de}
\partial^{\da\a} \partial^{\db\b}
\partial^{\dg\g} \partial^{\dd\d} \phi_{bcde}
R_{\a\b\g\d}  
\\
%\langle\,V^- \, V^- \, \chi^+ \,\psi^+  \,\rangle  && 
\rule{0pt}{.4cm}
\< 6\,6\,3\,1| \,\circ&& 
- \<12\> \<1|3|4] \<2|3|4] s_{23} [34]
&&
\frac{2}{2} F_{\da\db}^{bc}  
\partial_\m F_{\hspace{.18cm}\dg}^{\da\hspace{.12cm}de}  
\partial^\m \partial^{\db\b} \partial^{\dg\g} 
\chi^\a_{bcd}
\psi_{\a\b\g\,e}  
\\
%\langle\, V^- \, V^- \, V^+ \, V^+  \,\rangle  && 
\rule{0pt}{.4cm}
\< 6\,6\,2\,2| \,\circ&& 
\< 12\>^2 s_{23}^2 [34]^2
&&
\tfrac{2^2}{2^3} F_{\da\db}^{bc} 
\partial_\m \partial_\n F^{\da\db\,de}
\partial^\m \partial^\n
F^{\a\b}_{bc}  
F_{\a\b\,de}  
\\
\rule{0pt}{.4cm}
&& 
\< 12\>^2 s_{23} s_{24} [34]^2
&&
\tfrac{2^2}{2}F_{\da\db}^{bc} 
\partial_\m \partial_\n F^{\da\db\,de}
\partial^\m 
F^{\a\b}_{bd}  
\partial^\n F_{\a\b\,ce} 
\\
%\langle\, V^- \, \chi^- \, \chi^- \, +  \,\rangle  && 
\rule{0pt}{.4cm}
\< 6\,5\,5\,0| &&  
\<1|3|4]^2  \<2|3|4] \<3|2|4] 
&&
 \tfrac{1}{3!^2 2^2} \e_{bcdefghi}
 F_{\da\db}^{bc}   \partial^{\dd\d} \chi_{\dg}^{def}
\partial^{\da\a}  \partial^{\db\b}   \partial^{\dg\g}  
\chi_{\dd}^{ghi}
R_{\a\b\g\d}
\\
%\langle\, V^- \, \chi^- \, \phi \,\psi^+  \,\rangle  && 
\rule{0pt}{.4cm}
\< 6\,5\,4\,1| && 
\<1|3|4]^2 \<2|3|4] s_{24}
&&
\tfrac{2\rmi}{3!}F_{\da\db}^{bc}  
\partial_{\m}  \chi_{\dg}^{def} 
\partial^{\da\a}  \partial^{\db\b}  
\partial^{\dg\g}  \phi_{bdef}\,
\partial^{\m}   \psi_{\a\b\g\,c}
\\
\rule{0pt}{.4cm}
&& 
-\<1|3|4]^2 \<2|3|4] s_{23}
&&
-\tfrac{2\rmi}{2^2}F_{\da\db}^{bc}  
\partial_{\m}  \chi_{\dg}^{def} 
\partial^{\m}  \partial^{\da\a}  \partial^{\db\b}  
\partial^{\dg\g}  \phi_{bcde} 
\psi_{\a\b\g\,f}
\\
%\langle\, V^- \, \chi^- \, \chi^+ \,V^+  \,\rangle  && 
\rule{0pt}{.4cm}
\< 6\,5\,3\,2| && 
\<12\> \<1|3|4]   s_{23}^2 [34]
&&
\tfrac{2^2\rmi}{2^2}
F_{\da\db}^{bc}  
\partial_{\m} \partial_{\n}   \chi^{\da\,def}
 \partial^{\m}  \partial^{\n}  \partial^{\db\b} 
  \chi^\a_{bcd} 
F_{\a\b\,ef}
\\
\rule{0pt}{.4cm}
&& 
- \<12\> \<1|3|4]   s_{23} s_{24} [34]
&&
-\tfrac{2^2\rmi}{2}
F_{\da\db}^{bc}  
\partial_{\m} \partial_{\n}   \chi^{\da\,def}
 \partial^{\m}    \partial^{\db\b} 
  \partial^{\n} \chi^\a_{bde} 
F_{\a\b\,cf}
\\
\rule{0pt}{.4cm}
&& 
- \<12\> \<1|3|4]   s_{24}^2 [34]
&&
-\tfrac{2^2\rmi}{3!2}
F_{\da\db}^{bc}  
\partial_{\m} \partial_{\n}   \chi^{\da\,def}
 \partial^{\db\b} 
  \chi^\a_{def} 
 \partial^{\m}  \partial^{\n} F_{\a\b\,bc}
\end{array}
\ee

\[
\begin{array}{lclcl}
\< 6\,4\,4\,2| && 
\<1|3|4]^2 s_{24}^2
&&
-\tfrac{2^2}{4!2^2}F_{\da\db}^{bc}
\partial_{\m} \partial_{\n} \phi^{defg}
\partial^{\da\a}   \partial^{\db\b} 
\phi_{defg}
\partial^{\m} \partial^{\n} F_{\a\b\,bc}
\\
\rule{0pt}{.4cm}
&& 
\<1|3|4]^2 s_{23} s_{24}
&&
-\tfrac{2^2}{3!2}
F_{\da\db}^{bc}
\partial_{\m} \partial_{\n} \phi^{defg}
\partial^{\m} \partial^{\da\a}   \partial^{\db\b} 
\phi_{bdef}
\partial^{\n}  F_{\a\b\,cg}
\\
\rule{0pt}{.4cm}
&& 
\<1|3|4]^2 s_{23}^2
&&
-\tfrac{2^2}{2^4}
F_{\da\db}^{bc}
\partial_{\m} \partial_{\n} \phi^{defg}
\partial^{\m} \partial^{\n} \partial^{\da\a}   \partial^{\db\b} 
\phi_{bcde}
F_{\a\b\,fg}
\\
\rule{0pt}{.4cm}
\< 6\,4\,3\,3|\,\circ && 
-\<1|2|3]  \< 1|2|4] s_{24}^2
&&
-\tfrac{2^2}{3!2^2}
F_{\da\db}^{bc}
\partial_{\m}   \partial_{\n}  \partial^{\da\a}   \partial^{\db\b}  \phi^{defg}
  \chi_{\a\,def}
  \partial^{\m} \partial^{\n}   \chi_{\b\,gbc}
\\
\rule{0pt}{.4cm}
&& 
\<1|2|3]  \< 1|2|4] s_{23} s_{24}
&&
\tfrac{2^2}{2^2} 
F_{\da\db}^{bc}
\partial_{\m}   \partial_{\n}  \partial^{\da\a}   \partial^{\db\b}  \phi^{defg}
\partial^{\m} \chi_{\a\,bde}
   \partial^{\n}    \chi_{\b\,cfg}
\\
%\langle\,\chi^- \, \chi^- \, \chi^- \,\psi^+  \,\rangle  && 
\rule{0pt}{.4cm}
\< 5\,5\,5\,1| \,\circ&&
\<1|2|4] \<2|3|4] \<3|1|4] s_{23}
&&
-\tfrac{2\rmi}{3!^2\,6 } \e_{bcdefghi}
\partial^{\dg\g}  \chi_{\da}^{bcd} 
\partial_\m \partial^{\da\a}  \chi_{\db}^{efg} 
\partial^\m  \partial^{\db\b}  \chi_{\dg}^{hij} 
 \psi_{\a\b\g\,j}  
\\
%\langle\,\chi^- \, \chi^- \, \phi \,V^+  \,\rangle  && 
\rule{0pt}{.4cm}
\< 5\,5\,4\,2|\,\circ &&
- \<1|3|4] \<2|3|4] s_{23}^2
&&
\tfrac{2^2}{3!2^2} 
\chi_{\da}^{bcd} \partial_{\m} \partial_{\n}  \chi_{\db}^{efg}
\partial^{\m} \partial^{\n} \partial^{\da\a} \partial^{\db\b} 
\phi_{bcde}
F_{\a\b\,fg}
\\
\rule{0pt}{.4cm}
&&
\<1|3|4] \<2|3|4] s_{23} s_{24}
&&
-\tfrac{2^2}{2^2} \chi_{\da}^{bcd} \partial_{\m} \partial_{\n}  \chi_{\db}^{efg}
\partial^{\da\a} \partial^{\db\b} 
\partial^{\m} \phi_{bcef}
\partial^{\n}  F_{\a\b\,dg}
\\
%\langle\,\chi^- \, \chi^- \, \chi^+ \,\chi^+  \,\rangle  && 
\rule{0pt}{.4cm}
\< 5\,5\,3\,3|\circ\circ &&
 \<12\>   s_{23}^3 [34] 
&&
-\tfrac{2^3}{3!^2} 
\chi_{\da}^{bcd}\,
 \partial_{\m} \partial_{\n} \partial_\r \chi^{\da\,efg}
\partial^{\m} \partial^{\n}   \partial^\r   \chi^\a_{bcd}
\chi_{\a\,efg}
\\
\rule{0pt}{.4cm}
&&
 \<12\>   s_{23}^2 s_{24} [34] 
&&
-\tfrac{2^3}{2^3}
\chi_{\da}^{bcd}\,
 \partial_{\m} \partial_{\n} \partial_\r \chi^{\da\,efg}
\partial^{\m} \partial^{\n}    \chi^\a_{bce}
\partial^\r  \chi_{\a\,dfg}
\\
%\langle\,\chi^- \, \phi \, \phi \,\chi^+  \,\rangle  && 
\rule{0pt}{.4cm}
\< 5\,4\,4\,3| && 
- \< 1|3|4] s_{23}^3 
&&
-\tfrac{2^3\rmi}{3!^2 2}
\chi_{\da}^{bcd}
\partial_{\m} \partial_{\n}  \partial_{\r} \phi^{efgh}
\partial^{\m} \partial^{\n}  \partial^{\r}
\partial^{\da\a} \phi_{bcde}
\chi_{\a\,fgh}
\\
\rule{0pt}{.4cm}
&&
\< 1|3|4]  s_{23}^2 s_{24}   
&&
\tfrac{2^3\rmi}{2^3 2}
\chi_{\da}^{bcd}
\partial_{\m} \partial_{\n}  \partial_{\r} \phi^{efgh}
\partial^{\m} \partial^{\n}  
\partial^{\da\a} \phi_{bcef}
\partial^{\r} \chi_{\a\,dgh}
\\
\rule{0pt}{.4cm}
&&
- \< 1|3|4]  s_{23} s_{24}^2   
&&
-\tfrac{2^3\rmi}{3!2^2}
\chi_{\da}^{bcd}
\partial_{\m} \partial_{\n}  \partial_{\r} \phi^{efgh}
\partial^{\m} 
\partial^{\da\a} \phi_{befg}
\partial^{\n}  \partial^{\r} \chi_{\a\,cdh}
\\
\rule{0pt}{.4cm}
&&
\< 1|3|4] s_{24}^3
&&
\tfrac{2^3\rmi}{4!3!2}
\chi_{\da}^{bcd}
\partial_{\m} \partial_{\n}  \partial_{\r} \phi^{efgh}
\partial^{\da\a} \phi_{efgh}
\partial^{\m} \partial^{\n}  \partial^{\r} \chi_{\a\,bcd}
\\
\rule{0pt}{.4cm}
%%%%%%%%%%%%%%%%%DAN%%%%%%%
\< 4\,4\,4\,4|\circ\circ &&
s_{13}^2 s_{24}^2
&&
\tfrac{2^4}{4!^2 2^3} 
 \partial_{\m} \partial_{\n}\phi^{bcde}
 \partial_{\r} \partial_{\s} \phi^{fghi}
 \partial^{\m} \partial^{\n}   \phi_{fghi}
\partial^{\r} \partial^{\s}\phi_{bcde}
\\
\rule{0pt}{.4cm}
&&
s_{14}s_{23}^2 s_{24}
&&
\tfrac{2^4}{3!^2 2^2}
 \partial_{\m}  \phi^{bcde}
\partial_{\n}\partial_{\r} \partial_{\s} \phi^{fghi}
 \partial^{\n} \partial^{\r}  \phi_{bcdf}
 \partial^{\m}\partial^{\s} \phi_{eghi}
\\
\rule{0pt}{.4cm}
&&
s_{14}s_{23}s_{13} s_{24}
&&
\tfrac{2^4}{2^7}
\pa_\m \pa_\s\phi^{bcde}
  \partial_{\n} \partial_{\r}  \phi^{fghi}
 \partial^{\m} \partial^{\r} \phi_{bcfg}
\partial^{\n} \partial^{\s}  \phi_{dehi}\,.\\
\end{array}
\]

%%%%%%%%%%%%%%%%%%%%%%

\section{The $D^{2k}R^4$ invariants}

Because of the simple relation \reef{d2ksamp} between their generating functions,  it
is very easy to obtain the component expansion of the general $D^{2k}R^4$ invariants from the tabulated results for $R^4$.
First we note that the symmetric polynomials $P_k(s,t,u)$ for the first three new cases are $P_2 =  
s^2+t^2+u^2,~~P_3= stu, ~~P_4 = (s^2+t^2+u^2)^2.$  The last one corresponds to a potential 7-loop divergence in $\cn=8$ SG.  

Suppose that $P(p_1, p_2,p_3,p_4)$ is \emph{any} polynomial in the external momenta and consider the generating function  $\cm_P \equiv P(p_i) \cm^0$  with $\cm^0$ given in \reef{r4samp}.   It should be clear that 
$\cm_P$  generates a set of amplitudes which satisfies linearized $\cn =8$ SUSY.  In our case the polynomials are constrained by Lorentz invariance and particle exchange properties to be the symmetric  $P_k(s,t,u)$.  It becomes an algebra problem rather than a physics problem to find these polynomials.  This problem was studied in \cite{Greenstu}.  For $k \ge 6$ there can be more than one independent $P_k$ for each value of $k$.

To see how to use this information, let's denote \emph{any} component operator in the tabulated expansion of $R^4 +\ldots$ very generically (and with all indices suppressed) by the quartic monomial $A(x)B(x)C(x)D(x)$.  
Then the effect of a factor  $s,~ t$ or $u$ in any polynomial $P_k$ is to change that monomial by applying  spacetime derivatives as follows:
\bea
{\rm factor} ~~ s=s_{12}  && \hspace{1.5cm}   2\pa_\mu A \pa^\m BCD\,,\\
 {\rm factor} ~~t=s_{13}  && \hspace{1.5cm}   2\pa_\mu A B\pa^\m CD\,,\\
 {\rm factor}~~u=s_{14} && \hspace{1.5cm}   2\pa_\mu A BC \pa^\m D\;.
  \eea
Higher order factors are easily included repeating the same basic rule.  For example the effect of a factor  $s^2$ in $P_k$ is the field monomial 
$4 (\pa_\m\pa_\n A)(\pa^\m\pa^\n B) CD$ and a factor $st$ gives
$4  (\pa_\m\pa_\n A)(\pa^\m B) (\pa^\n C) D$.   In this way the effect of all factors in the polynomial $P_k(s,t,u)$ can be obtained.   The important point is that  \emph{exactly the same derivatives are inserted in each of the
51 quartic monomials of the component expansion of $R^4 +..$}. This  procedure completely determines  the component expansion of $D^{2k}R^4 +..$\,.

\section{Discussion}

The principal result of our work is the detailed component form of the $R^4$ invariant in $\cn =8$ four-dimensional supergravity\footnote{Partial results in 10 and 11 dimensions were presented in \cite{peeters}.}.  This is expressed as the sum of 51 terms given in (\ref{LIST}).  It was constructed from the information in the simple
4-point MHV superamplitude \reef{mhvsamp}.  The construction guarantees linearized $\cn=8$ supersymmetry and manifest SU(8) R-symmetry.  Analogous component forms of the $D^{2k}R^4$ invariants can be obtained by applying $2k$  spacetime derivatives to the same set of 51 terms.  We cannot point to any immediate application of these results, but we hope that the display of the full component content is instructive.  

The gravitational part of $R^4 + \ldots$ was identified as the square of the Bel-Robinson tensor in \cite{dks} and a linearized $\cn =1$ SUSY completion was given there.  It is well known \cite{penrose} that the spinor form of the on-shell Bel-Robinson tensor is exactly $R_{\a\b\c\d} R_{\da\db\dg\dd}$.  Indeed its square is the purely gravitational term in our \reef{LIST}.

For  $\cn =8$ SG, the $R^4$ invariant was first presented \cite{K1981}   in 1981 using the (linearized) on-shell superspace formalism of \cite{brink}.  
This a superspace with 16 2-component $\theta_a$ plus 16 conjugate $\bar{\theta}^a,~ a=1,\ldots 8$.
The basic superfield is the self-dual fourth rank symmetric
$W_{abcd}= \e_{abcdefgh} \bar{W}^{efgh}/4!$\,. Not surprisingly the lowest component of the $\theta$ expansion of $W_{abcd}$ is the component scalar field $\phi_{abcd}$.  The invariant is expressed as an integral over a 16-dimensional subspace of the superspace, obtained using a special "proper basis"  in which only $\theta_a$ with $a=1,2,3,4$ and $\bar{\theta}^{a'}$ with $a'=5,6,7,8$ appear:
\be \lab{r4kallosh}
S \sim \int d^4x \,d\mu^{abcdefgh}d\bar{\mu}_{a'b'c'd'e'f'g'h'} \,W_{abcd}W_{efgh}\bar{W}^{a'b'c'd'}\bar{W}^{e'f'g'h'}\,.
\ee
The measure is an integral over the $8+8~ \theta$'s of the proper basis.   Only  SU(4)$\times$SU(4) symmetry is manifest in the construction, but properties of the $\theta$-expansion in the proper basis
ensure SU(8) invariance of the result \cite{K1981}.  A manifestly SU(8) invariant version of the counterterm has also been found \cite{hst1981}. The integrand  contains the product of  four $W_{abcd}$'s  (with full  index range 1-8).  It transforms in the symmetric product of four 70-dimensional irreps of SU(8),  which has dimension 232848.   The product is contracted with a 16 $d\theta$ measure in the same representation.  In \cite{hst1981} the invariant form was deemed "probably equivalent" to \reef{r4kallosh}. 

The  $R^4$ invariant has also been expressed  as \cite{drum2003} the 16-dimensional sub-integral  in on-shell harmonic superspace 
\be \lab{r4howe}
S \sim \int d\mu_{(4,4)} (W_{1234})^4\,.
\ee
The measure is given in \cite{drum2003}.  Very recently \cite{drum2010} the question of sub-superspace invariants for $\cn = 8$  was reanalyzed
using superconformal symmetry.  No new formula for the $R^4$ invariant was given.   The various superspace arguments outlined above all establish linearized $\cn = 8$ SUSY.

The 3-, 5-, and 6-loop invariants  $R^4, D^4R^4$, and $D^6 R^4$ all involve sub-superspace integrals.  However,  
the $D^{2k}R^4$ invariants for $k \ge 4$ are expressed as full superspace
integrals \cite{K1981,Lind1981}, whose integrands involve supercurvature and supertorsions.  Thus they can be expressed in terms of
geometric quantities.  In fact, the potential 7-loop $D^8 R^4$ invariant is just the total superspace volume \cite{Lind1981},\cite{Bossard}.  
 No such sharp distinction between the ranges $k <4$ and $k \ge 4$ occurs in the matrix element based construction \cite{efk} of $D^{2k}R^4$.  However, the results of \cite{efk} and \cite{befkms} clearly indicate that many independent candidate counterterms with $n \ge 5$ external particles 
 are available beginning at loop level $L=7$.
 \newpage

\subsection*{Acknowledgments}

We are happy to thank Henriette Elvang and Michael Kiermaier for useful discussions and suggestions.\\
ET is mainly supported by Istituto Nazionale di Fisica Nucleare (INFN) through a Bruno Rossi fellowship.   DZF's principal external support comes
from the NSF grant PHY-0967299.   Both DZF and ET are also supported 
 by funds of the U.S. Department of Energy  under the cooperative research agreement DE-FG02-05ER41360.

\appendix
 
 \section{Spinor conventions} \label{app spinors}

We collect useful formulas from Appendix A of \cite{bef}.  Conventions for 2-component spinors were deduced from the 4-component formalism using
\be
\g^\m =
\left(\begin{array}{cc}
0 & (\s^\m)_{\a\db} \\
(\bar{\s}^\m)^{\da\b} & 0
\end{array}\right)\,,
\hspace{1.5cm}
\s^\m = (1,\s^i)\,,
\hspace{1cm}
\bar{\s}^\m = (-1,\s^i)\;.
\ee
Note that $(\bar{\s}^\m)^{\da\b}  =  - \e^{\da\dg}\e^{\b\d}(\s^\m)_{\d\dg}$
with  $\e^{12}=\e_{12} =\e^{\dot{1}\dot{2}}= \e_{\dot{1}\dot{2}}=1\,.$  The bispinor forms of any 4-vector, such as a partial derivative $\pa_\m$ or momentum $p_\m$,   are
\be  \lab{bispin}
p_{\a\db} = p_\mu (\s^\m)_{\a\db}\,,  \hspace{2cm}  p^{\da\b} = p_\m  (\bar{\s}^\m)^{\da\b}\,.
\ee
For any null momentum $p_\m$,   (commuting) 2-component spinors  $\l_{\a}(p)$ and $\tl_{\da}(p)$,
which are related by complex conjugation $\l_\a = (\tl_\da)^\ast$, are defined as solutions of the Weyl equations
\be \lab{weyleq}
p^{\da\b} \,\l_\b =0\,,  \hspace{2cm} p_{\a\db}\,\tl^{\db}=0\,.
\ee 
Spinor indices are raised as $\l^a =\e^{\a\b}\l_\b, ~~\tl^\da =\e^{\da\db}\tl_\db$ and lowered as $\l_\a = \l^\b\e_{\b\a},~~\l_\da = \tl^{\db}\e_{\db\da}\,.$  One can  show that  $\tl^\da(p) \l^\b(p) = p^{\da\b}$, but $\l_\a(p)\tl_\db(p) = - p_{\a\db}$.
Spinor brackets are defined by
\be
\<pq\> = \tl_{\da}(p)\tl^{\da}(q)\,, \hspace{2cm}  [pq] = \l^\a(p) \l_{\a}(q)\,.
\ee
The particles in a scattering process are always numbered and we use the common notation  $\<i j\> = \<p_ip_j\>$, etc..  The angle-square bracket
\be \lab{angsq}
\<i|k|j] = \<i k\>[kj] =\tl_{\da}(p_i)\,\tl^\da(p_k) \l^\b(p_k)\, \l_{\b}(p_j) = \tl_{\da}(p_i)\,p_k^{\da\\b}\, \l_{\b}(p_j)\,
\ee
occurs frequently in our work.  Mandelstam invariants are given by
$\<ij\>[ij] = 2 p_i\cdot p_j = -s_{ij}$.

Let $\bar{\ve} Q$ denote the 4-component bilinear of a Majorana supercharge and SUSY parameter.  In these conventions this quantity is related to the chiral supercharges by
\be 
\bar{\ve}Q =  -\rmi (\tilde{\ve}_\da \tQ^\da - \ve^\a Q_\a) = - \rmi (Q + \tQ)\,.
\ee
The 4-component SUSY variation of a any field $\phi$ is then implemented
as
\be
\d_{\ve}\phi = \rmi[  \bar{\ve}Q, \phi] = [Q + \tQ, \phi]\,. 
\ee
 Chiral SUSY variations
$\d \phi$ and $\tilde{\d}\phi$ are therefore implemented without prefactor $\rmi$.  The sum $Q +\tQ$ is anti-hermitiean.

\section{Examples of SUSY checks}
\label{app susy}

In this appendix we outline the SUSY checks which confirm that the first four terms of the component expansion of the $R^4 +\ldots$ invariant are correct.  First, some general considerations.

We consider $Q^a$ variations of the monomials in
the component expansion.  To each monomial there is a partition of 16 whose entries encode  the number of upper SU(8) indices carried by each of the four fields.   Since the $Q^a$ variation adds one upper index the number of independent quartic monomials in the variation is the number of partitions of 17 of length 4 and maximal summand 8.   There are 31 such partitions.  Since a $\tQ_a$ variation subtracts an upper index, the number of quartic monomials it generates is equal to the number of partitions of 15  
of length 4 and maximal summand 8.   The CPT conjugation properties of
$\cn = 8$ SG imply that this number must also be 31 which it is.

Let's begin the technical work by defining the variations of the first four terms in the \reef{LIST}.   In the notation of partitions, and using $(\d^aS)$ to indicate the spacetime integrals of their variations, we have
\bea  \lab{4vars}
\d^a \int \mathcal{L}^{(0088)} &=& (\d^aS)^{(0188)}_1\,, \nonumber \\
\d^a \int \mathcal{L}^{(0178)} & =&
(\d^a S)^{(0188)}_2+ (\d^a S)^{(0278)}_2+ (\d^a S)^{(1178)}_2\,,  \nonumber\\ 
\d^a \int \mathcal{L}^{(0268)} & =&
(\d^a S)^{(0278)}_3+ (\d^a S)^{(0368)}_3+ (\d^a S)^{(1268)}_3\,,   \nonumber\\
\d^a \int \mathcal{L}^{(0277)} & =&
(\d^a S)^{(0278)}_4+ (\d^a S)^{(0377)}_4+ (\d^a S)^{(1277)}_4\,.   
\eea

The only Lagrangian monomials whose variation populates the $(0188)$ partition are $\cl^{(0088)}$ and $\cl^{(0178)}$,
 so SUSY requires that $(\d^a S)^{(0188)}_1 +(\d^a S)^{(0188)}_2 =0 $.   
To show that the necessary cancellation occurs, we use the $\d^a R_{\da\db\dg\dd}$ variation from \reef{trafos} to write
\be
\label{susy 0088}
(\d^a S)^{(0188)}_1=
-\frac{\rmi}{2} \int \ve^\s \pa_{\s\da} \psi^a_{\db\dg\dd} R^{\da\db\dg\dd} 
R^{\a\b\g\d} R_{\a\b\g\d}\,.
\ee
We use the symmetric derivative property to exchange indices $\da \lra \dd$,  integrate by parts, and exchange $\s \lra \d$ to obtain
\be
\label{susy 0088 bis}
(\d^a S)^{(0188)}_1= \rmi \int \ve^\s  \psi^a_{\da\db\dg} R^{\da\db\dg\dd} R^{\a\b\g\d}
\pa_{\d\dd} R_{\s\a\b\g}\;.
\ee
Next we consider 
\be  \lab{s0178}
(\d^a S)^{(0188)}_2   =\,\rmi \int R_{\da\db\dg\dd} \psi^{\da\db\dg\,a}
\ve_\s\pa^{\dd\d}  R^{\s\a\b\g} R_{\a\b\g\d} \,.
\ee
The two expressions differ in the position of indices.  After raising indices with careful attention to the conventions of Appendix \ref{app spinors} and moving $\ve^\s$ to the left of  $ \psi_{\da\db\dg}^a$, we find that the  variations \reef{susy 0088 bis} and \reef{s0178} cancel.

In the next set of calculations,  which involve the cancellation of three different contributions to $(\d^aS)^{(0278)}$,  it is unfeasible to work directly
with the variations as we did above.  Instead we devised a method to convert the information in the variations into on-shell matrix elements.
Standard manipulations in the spinor-helicity formalism can then be used to  establish the required cancellations. 
 To  illustrate this method let us apply it to the simple case just treated above.
 
  We consider an outgoing state with fixed SU(8) indices which  "communicates"  to  $(\d^aS)^{(0188)}_1$ and $ (\d^aS)^{(0188)}_2$.  Thus we are led to the 
 matrix element of the variation (\ref{susy 0088}) which we calculate via Wick contractions:
\be
\label{susymat 0088}
\< \textrm{out} | A_8(1) A^+(2) A_-(3) A_-(4)  \,
\frac{-\rmi}{2} \int \ve^\s \pa_{\s\da} \psi^a_{\db\dg\dd} R^{\da\db\dg\dd} R^{\a\b\g\d} R_{\a\b\g\d} 
| \textrm{in} \> =
- \< 12\>^4 [34]^3 [\ve 1] [34]\,.
\ee
Notice that the two Wick contractions provide the same result for the matrix element, which is invariant under $3 \leftrightarrow 4$ exchange.
Then, we compute the same matrix element of the term (\ref{s0178}): 
\bea
\label{susymat 0178}
\< \textrm{out} | A_8(1) A^+(2) A_-(3) A_-(4)\; \rmi
\int R_{\da\db\dg\dd} \psi^{\da\db\dg\,a}
\pa^{\dd\d} \ve_\s R^{\s\a\b\g} R_{\a\b\g\d} 
| \textrm{in} \> \,= & & \\
& & \hspace{-3cm} =\,
- \< 12\>^4 [34]^3 [\ve 3] [41]- \< 12\>^4 [34]^3 [\ve 4] [13]\,.
\nonumber
\eea
This time the two Wick contractions give different matrix elements. It is gratifying that the sum of  (\ref{susymat 0088}) and (\ref{susymat 0178}) vanishes because of the Schouten identity
\be
- \< 12\>^4 [34]^3 \Big( [\ve 1] [34]+ [\ve 3] [41]+[\ve 4] [13]\Big) =0\;.
\ee
Thus, the SUSY check confirms the relative factor between these two Lagrangian counterterms $\mathcal{L} ^{(0088)}$ and $\mathcal{L} ^{(0178)}$ which was found through the method explained in the section \ref{section R4 inv}.

%%%%%%%%%%%%%%%%%%%%
The most common situation that occurs in a check of component SUSY is that three different Lagrangian monomials contribute to each fixed independent variation $(\d^aS)$.   From \reef{4vars}, we see that this is the case for
$(\d^aS)^{(0278)}$.  Therefore, our
 next task is to compute the variations $(\d^aS)^{(0278)}_i$ for $i=2,3,4$
and show by the matrix element method that their sum vanishes.  The variations are computed from the transformation rules in \reef{trafos}:
\bea \lab{3vars}
(\d^aS)^{(0278)}_2 &=&-\int R_{\da\db\dg\dd} \ve_\s \pa^{\da\s} F^{\db\dg\,ab} \pa^{\dd\d} \psi^{\a\b\g}_b R_{\a\b\g\d}\,,\\
(\d^aS)^{(0278)}_3 &=&-\int R_{\da\db\dg\dd} F^{\da\db\,ab} 
\ve_\s \pa^{\dg\g} \pa^{\dd\d} \psi^{\s\a\b}_b R_{\a\b\g\d}\,,\\
(\d^aS)^{(0278)}_4 &=&\int R_{\da\db\dg\dd} \pa^{\dg\g} \pa^{\dd\d} F^{\da\db\,ab}
\psi^{\a\b}_{\hspace{.32cm}\g\,b} \ve_\s R^\s_{\hspace{.14cm}\a\b\d}\,.
\eea
Next we use Wick contractions to compute the three matrix elements
\bea\lab{3me}
\< \textrm{out} | A^+(1) A_{12}(2) A^2(3) A_-(4) (\d^aS)^{(0278)}_2 | \textrm{in} \>  &=& \<12\>^3 \<13\> [2\ve] [34]^4\,,\\
\< \textrm{out} | A^+(1) A_{12}(2) A^2(3) A_-(4) (\d^aS)^{(0278)}_3 | \textrm{in} \>  &=& \<12\>^2 \<13\>^2 [3\ve] [34]^4\,,\\
\< \textrm{out} | A^+(1) A_{12}(2) A^2(3) A_-(4) (\d^aS)^{(0278)}_4 | \textrm{in} \>  &=& \<12\>^2 \<13\> \<14\>  [4\ve] [34]^4\,.
\eea
Finally we can check that the sum of these three matrix elements vanishes by momentum conservation, viz.
\be
\<12\>^2 \<13\> \Big( \<12\>[2\ve] +\<13\>[3\ve] +\<14\>[4\ve] \Big) [34]^4 =0\,.
\ee
An analogous SUSY check confirms the detailed monomials given in \reef{LIST} for the sectors (8710), (7720) and (7711).

%%%%%%%%%%%%%%%%

\end{document}